\documentclass[preprint2,twoside]{hwo}

\usepackage{graphicx}
\usepackage{chemformula}
\usepackage{upgreek}
\usepackage{makecell}

\DeclareRobustCommand{\ion}[2]{\textup{#1\,\textsc{\lowercase{#2}}}}
\newcommand{\Lyalpha}{Ly-$\alpha$}
\newcommand{\helium}{\ion{He}{I}}
\newcommand{\MgII} {\ion{Mg}{ii}}

\newcommand{\FeII} {\ion{Fe}{II}}

\newcommand{\SiII} {\ion{Si}{II}}

\input{hwo.h}

\begin{document}
\renewcommand{\thefootnote}{\fnsymbol{footnote}}

\title{\textbf{\LARGE High-resolution Ultraviolet-to-nearinfrared Characterization of Exoplanet Atmospheres}}
\author{\textbf{\large Patricio E. Cubillos$^{1,2,}$\footnote{\href{patricio.cubillos@oeaw.ac.at}{patricio.cubillos@oeaw.ac.at}}, Matteo Brogi$^{2,3}$, Antonio García Muñoz$^{4}$, Luca Fossati$^{1}$, Sudeshna Boro Saikia$^{5}$, Vincent Bourrier$^{6}$, Jose A. Caballero$^{7}$, Juan Cabrera$^{8}$, Andrea Chiavassa$^{9}$, Andrzej Fludra$^{10}$, Leonardos Gkouvelis$^{11}$, John Lee Grenfell$^{8}$, Manuel Guedel$^{5}$, Alvaro Labiano$^{12}$, Monika Lendl$^{6}$, Donna Rodgers-Lee$^{13}$, Arnaud Salvador$^{8}$, Ilane Schroetter$^{14}$, Antoine Strugarek$^{4}$, Benjamin Taysum$^{8}$, Aline Vidotto$^{15}$, and Thomas G. Wilson$^{16}$}}
\affil{$^1$\small\it Space Research Institute, Austrian Academy of Sciences, Schmiedlstrasse 6, A-8042, Graz, Austria}
\affil{$^2$\small\it INAF -- Osservatorio Astrofisico di Torino, Via Osservatorio 20, 10025 Pino Torinese, Italy}
\affil{$^3$\small\it Dipartimento di Fisica, Universitá degli Studi di Torino, via Pietro Giuria 1, I-10125, Torino, Italy}
\affil{$^4$\small\it Université Paris-Saclay, Université Paris Cité, CEA, CNRS, AIM, 91191, Gif-sur-Yvette, France} 
\affil{$^5$\small\it  Department of Astrophysics, University of Vienna, Türkenschanzstraße 17, 1180 Vienna, Austria}
\affil{$^6$\small\it Observatoire de Genève, Département d'Astronomie, Université de Genève, Chemin Pegasi 51b, 1290 Versoix, Switzerland}
\affil{$^7$\small\it  Centro de Astrobiología (CSIC-INTA), ESAC Campus, Camino bajo del castillo s/n, 28692 Villanueva de la Cañada, Madrid, Spain}

\affil{$^8$\small\it Deutsches Zentrum für Luft- und Raumfahrt (DLR), Institut für Planetenforschung, Rutherfordstrasse 2, 12489 Berlin, Germany}
\affil{$^9$\small\it Université Côte d'Azur, Observatoire de la Côte d'Azur, CNRS, Lagrange, CS 34229, Nice, France}
\affil{$^{10}$\small\it RAL Space, UKRI STFC Rutherford Appleton Laboratory, Chilton, UK}
\affil{$^{11}$\small\it Ludwig Maximilian University, Faculty of Physics, University Observatory, Scheinerstrasse 1, Munich D-81679, Germany}
\affil{$^{12}$\small\it Telespazio UK for the European Space Agency, ESAC, Camino Bajo del Castillo s/n, 28692 Villanueva de la Cañada, Spain}
\affil{$^{13}$\small\it Dublin Institute for Advanced Studies, 31 Fitzwilliam Place, Dublin D02 XF86, Ireland}
\affil{$^{14}$\small\it Institut de Recherche en Astrophysique et Planétologie, Université de Toulouse, CNRS, CNES, Toulouse, France}
\affil{$^{15}$\small\it Leiden Observatory, Leiden University, P.O. Box 9513, NL-2300 RA Leiden, the Netherlands}
\affil{$^{16}$\small\it Department of Physics, University of Warwick, Gibbet Hill Road, Coventry CV4 7AL, UK}

\author{\footnotesize{\bf Endorsed by:}
Emanuele Bertone (Instituto Nacional de Astrofísica, Óptica y Electrónica),
Jayne Birkby (University of Oxford),
Abby Boehm (Cornell University),
Jean-Claude Bouret (Laboratoire d'Astrophysique de Marseille),
Adam Burgasser (UC San Diego),
Aarynn Carter (STScI),
Vincent Esposito (Chapman University),
Ryan Fortenberry (University of Mississippi),
Caleb Harada (UC Berkeley),
Finnegan Keller (Arizona State University),
James Kirk (Imperial College London),
Adam Langeveld (Johns Hopkins University),
Eunjeong Lee (EisKosmos CROASAEN),
Mercedes López-Morales (Space Telescope Science Institute),
Nataliea Lowson (University of Delaware),
Evelyn Macdonald (University of Vienna),
Stanimir Metchev (University of Western Ontario),
Drew Miles (California Institute of Technology),
David Montes (Universidad Complutense de Madrid),
Coralie Neiner (LIRA, Paris Observatory),
Seth Redfield (Wesleyan University),
Blair Russell (Chapman University),
Farid Salama (NASA Ames Research Center),
Gaetano Scandariato (INAF),
Vincent Van Eylen (UCL),
Peter Wheatley (University of Warwick)
}



\begin{abstract}
The Habitable Worlds Observatory (HWO) offers a unique opportunity to
revolutionize our understanding of planetary formation and evolution.
The goal of this Science Case Development Document (SCDD) is to
investigate the physical and chemical processes that shape the
composition and atmospheric mass loss in exoplanets.  We review the
key observables currently known as diagnostics of mass loss via
transit observations, i.e., absorption lines of escaping hydrogen
({\Lyalpha}), helium, and metals (Fe, Mg, C, O).  We also explore the
challenges to infer planetary formation processes based on atmospheric
composition characterization.
HWO could enable a broad, continuous coverage from far-ultraviolet to
near-infrared spectroscopy ($\sim$100--1600~nm) at high resolution ($R
\gtrsim 60,000$), which is essential to make these measurements,
disentangle their planetary origin from stellar activity, and
ultimately, contextualize the escape rates by simultaneously
characterizing the composition, cloud predominance, and thermal
structure of exoplanet atmospheres.
\\

\end{abstract}

\vspace{2cm}

\section{Science Goal}

The fundamental question we aim to address in this HWO science case
is: \textbf{\textit{How do exoplanet systems form and evolve into their current
observed states?}} \\

\begin{figure*}[t]
\centering
\includegraphics[width=0.91\linewidth, clip]{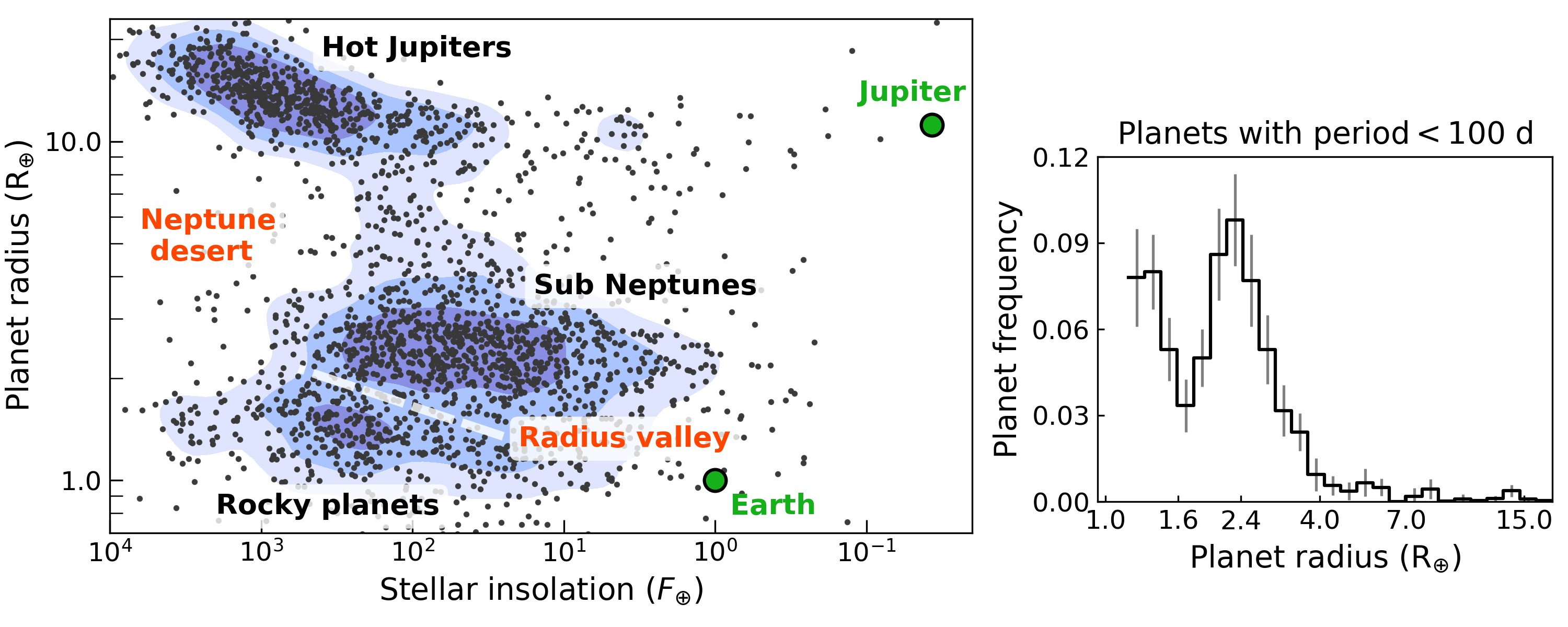}
\caption{Left panel: amount of stellar flux received as a function
  of planetary radius for the sample of known exoplanets (black
  dots). $F_\oplus$ is the Earth's incident irradiation from the Sun. The
  large blue dots show the position of Earth and Jupiter. The
  black and green labels denote the location where exoplanets are
  common (Jupiter, sub-Neptune, and rocky planets) and scarce
  (Neptune desert; radius valley, right panel), respectively.
  Right panel: radius valley distribution for planets in
  short-period orbits \citep{FultonEtal2017ajRadiusGap}. The
  structure in this distribution is believed to be the outcome of a combination
  of planet formation and planet evolution processes.}
\label{fig:fig1}
\end{figure*}

The thousands of extra-solar planets discovered to date have revealed
a staggering diversity of physical properties and environments in our
galaxy. Some bulk properties of this population (e.g., planetary
radius, mass, and incident stellar irradiation) have revealed distinct regions
in parameter space, some densely populated with exoplanets (e.g.,
forming groups of Jupiter, Neptune, or rocky-like planets) and others
sparsely populated. This distribution is thought to be the outcome of
planet formation and evolution processes, which can be understood by
studying the structure, composition, and loss of exoplanetary
atmospheres.

However, bulk properties alone are not sufficient to
unveil the true nature of an exoplanet. Formation and evolution
may create planets with similar masses and radii, but with strikingly
different atmospheric compositions. For example, a sub-Neptune mass
planet that retained its primordial \ch{H2}/He atmosphere may have a
similar radius as that of a super-Earth sized planet that has lost its
primary atmosphere due to intense atmospheric escape. Furthermore, the
planetary environment created by the star (radiation, particles,
stellar wind, magnetic fields) matters for atmospheres, linking the
atmospheric evolution of a planet to the properties of its host star over time. Only by
characterizing the composition of exoplanets can we unveil their true
nature and the physical processes that formed them. This will enable
us to place the Solar System and our own planet Earth in the broader
context of the known universe, which will bring us to uncover the
physical conditions that led to the emergence of life on Earth and to
identify other planets that have developed habitable conditions.

\section{Science Objective}

\subsection{Atmospheric characterization as a path to study planetary formation and
evolution processes}

Figure \ref{fig:fig1} shows the distribution of planetary radii versus stellar
insolation. Planet formation and evolution processes are believed to
be the main drivers shaping this distribution. For instance, planets
with short orbital periods experience intense high-energy irradiation
from their host stars, leading to significant atmospheric loss (also
known as atmospheric escape). This process strips their primordial \ch{H2}/He
envelopes, reducing their radii and masses over time
\citep[e.g.,][]{KurokawaNakamoto2014apjMassLossHotJupiters,
  OwenLai2018mnrasPhotoevaporationMigrationDesert}. As a result, there
is a scarcity of short-period planets with sizes smaller than Jupiter
and larger than Earth that is known as the ``Neptune desert''
\citep{MazehEtal2016aaNeptuneDesert,
  LundkvistEtal2016natcoSuperEarthsEvaporation}.  Only the most
massive planets are expected to retain their primary atmospheres,
delineating the boundary between the Neptune desert and the ``Hot Jupiters'',
although studies indicate that atmospheric escape alone cannot fully
explain the location of this boundary
\citep{VissapragadaEtal2022ajDesertPhotoevaporation}; other
factors, such as tidal disruption during a high-eccentricity migration
phase of planet formation likely also play a part
\citep[e.g.,][]{BeaugeNesvorny2013apjPeriodRadiusTrends,
  CastroGonzalezEtal2024aaNeptuneSavanna}.

\begin{figure*}[t!]
\centering
\includegraphics[width=\linewidth, clip, trim=2 0 0 0]{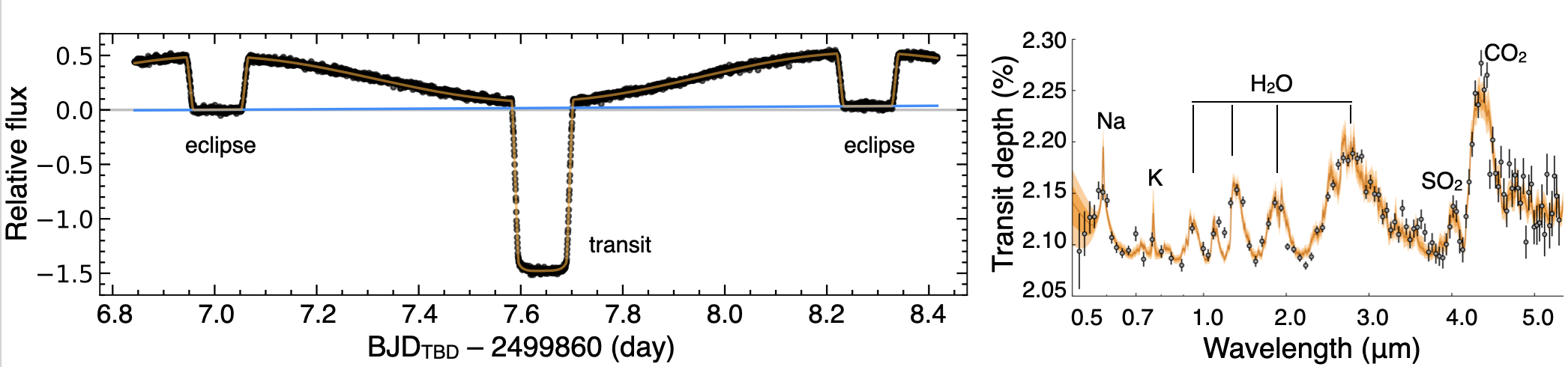}
\caption{Left panel: time-series observation of the transiting
  exoplanet WASP-121b during a full planet orbit with JWST/NIRISS
  \citep[44~h,][]{Mikal-EvansEtal2023apjWASP121bPhaseCurve}. The blue line
  shows the stellar baseline flux. Right panel: spectroscopic transit
  signature of the exoplanet WASP-39b combining four JWST observations
  \citep[NIRISS, NIRCam, NIRSpec, and MIRI;][]{CarterEtal2024natasDataSynthesisWASP39b}. The orange curve
  and shaded areas show a best-fitting theoretical model. The labels
  denote the atmospheric species responsible for the detected
  absorption features.}
\label{fig:fig2}
\end{figure*}

The other prominent feature in this population is a blended bimodal
distribution of small short-period planets separated by a ``Radius
valley'' at around 1.8 $R_\oplus$ \citep{FultonEtal2017ajRadiusGap,
  FultonPertigura2018ajGaiaCKS, VanEylenEtal2018mnrasRadiusValley, HoVanEylen2023mnrasDeepRadiusValley}. This is thought to be the mixture of
two distinct populations: the smaller ``super-Earth'' planets 
mainly composed of a rocky core, and the larger ``sub-Neptune'' planets
with can either host a massive \ch{H2}/He atmosphere or have a water-rich
interior and atmosphere \citep{AdamsEtal2008apjMassRadiusRelationship,
  RogersSeager2010apjInteriorDegeneracies,
  ValenciaEtal2013apjSubNeptunesComposition,
  VenturiniEtal2020aaRadiusValleyFormation}. Most intriguing for
placing the Solar System in the broader context, neither of these
types of objects have analogues in the solar system.

Fortunately, formation and escape processes not only affect the bulk
properties of exoplanets, but also directly shape the composition of
their atmospheres.  Characterizing the atmospheres of individual
targets is thus key to unveiling the processes that formed
them. Initial studies linking composition to planet formation
suggested that we can infer where a planet formed by comparing its
composition to that of the protoplanetary disk from which it
originated. Protoplanetary disk models predict that volatile gasses
(e.g., \ch{H2O}, \ch{CO2}, and CO) freeze out at different distances from the
central star, leading to a radial gradient in the C/O ratio of gases
and solids \citep{ObergEtal2011apjCOsnowlines}. By measuring a
planet's atmospheric C/O ratio, we can infer that it formed in the
region of the disk that most closely resembles the observed C/O.
However, later studies indicate that planet formation is far more
complex than that. A variety of formation mechanisms must be
considered, including planet migration, accretion history, and the
chemical structure of a dynamically evolving protoplanetary disk. The
combination and interplay of these processes can lead to a wide
diversity of planetary masses, orbital architectures, and atmospheric
compositions \citep[e.g.,][]{Madhusudhan2019aaraExoplanetAtmospheres,
  ZhuDong2021araaExoplanetStatistics,
  PacettiEtal2022apjDiskPlanetsChemicalDiversity,
  EistrupHenning2022aaPlanetFormationIcesPebbles}.

Each formation or evolution mechanism produces distinct and
potentially observable signatures in the final atmospheric composition
of a planet. For instance, in scenarios where planetary growth is
dominated by pebble accretion (i.e., accretion of small particles
composed of ice and dust), the resulting atmospheres are expected to
be enriched in volatile elements such as carbon and oxygen
\citep{SchneiderBitsch2021aaVolatilesRefractories,
  Crossfield2023apjlSulfurAccretionhistory}. Conversely, if the
dominant growth mechanism is planetesimal accretion (i.e., the
accretion of larger particles with rocky or metallic cores), the
resulting atmospheres should be enhanced in refractory elements such
as sodium, potassium, or sulfur
\citep{TurriniEtal2021apjFormationHistoryCNOS,
  PacettiEtal2022apjDiskPlanetsChemicalDiversity,
  Crossfield2023apjlSulfurAccretionhistory}. Furthermore, the total
atmospheric metallicity is expected to differ between these scenarios:
dominant pebble accretion can result in metallicities exceeding
10$\times$ solar, while planetesimal accretion is expected to yield
metallicities ranging from solar to a few times solar.

While current observatories have enabled the detection of many
observable features of interest in exoplanet atmospheres \citep[see
  review by][]{KemptonKnutson2024rmgTransitingExoplanets}, most
formation mechanisms remain poorly constrained due to the degenerate
nature of observable properties that they produce
\citep[e.g.,][]{MolliereEtal2022apjCompositionPlanetFormation}.
It is still unclear what the ideal observable(s) are to link formation
scenarios and exoplanet atmospheres---whether they be multiple
elemental abundance ratios, refractory budgets, or something else
\citep{FeinsteinEtal2025arxivLinkFormationAtmospheres}.  Disentangling
the various formation and evolutionary pathways requires observing
atmospheres of statistically significant samples of planets
\citep[e.g.,][]{PenzlinEtal2024mnrasFormationMigrationImpacts}, and a
comprehensive characterization of the atmospheric composition of these
planets, including the abundances of multiple key volatile and
refractory species (e.g., \ch{H2O}, CO, \ch{CO2}, \ch{CH4}, \ch{H2S},
SO, \ch{SO2}, Na, K), condensates (e.g., cloud and haze particles),
and escaping species (e.g., H, He, and metals); as well as the
monitoring of their host stars' activity. Such measurements demand
observations with high spectral resolution, high signal-to-noise
ratios, and broad spectral coverage---ideally achieved simultaneously.

\begin{figure*}[t!]
\centering
\includegraphics[width=\linewidth, clip]{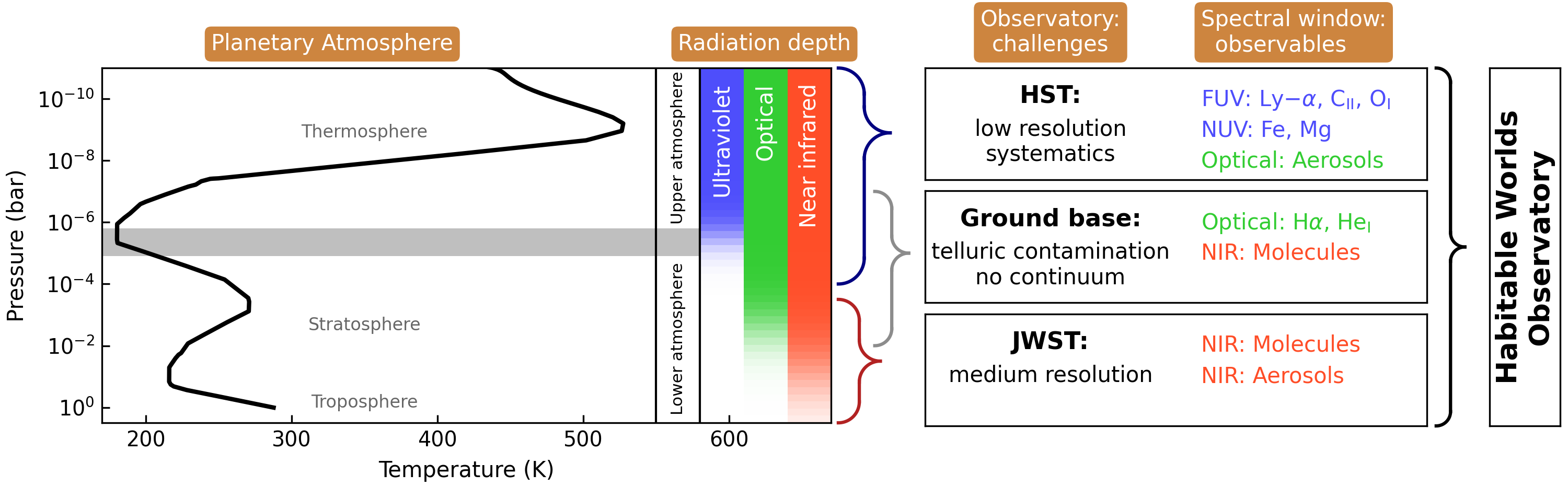}
\caption{Schematic view of a planetary atmosphere showing the Earth's
  pressure-temperature profile. The gray shaded region highlights the
  transition between the lower and upper atmosphere around the $10^{-6}$
  bar level. The middle panel illustrates the typical radiation
  penetration depth across different spectral bands—from ultraviolet
  (UV) to optical and near-infrared (NIR)---where different molecular
  and atomic species can be probed. The right panel summarizes
  observational challenges and spectral sensitivities of key
  observatories. The Hubble Space Telescope (HST) enables UV and
  optical access but is limited by low/medium resolution and presents
  significant instrumental systematics. Ground-based observatories
  face telluric contamination and lack of continuum access but can
  probe optical and NIR features. The James Webb Space Telescope
  (JWST) offers medium resolution coverage in the near and mid
  infrared, sensitive to molecular and aerosol signatures, but has no
  access to the UV.  The Habitable Worlds Observatory is envisioned to
  expand multi-wavelength access and enhance the detection of a wide
  array of atmospheric species across different pressure layers.  }
\label{fig:fig3}
\end{figure*}

These atmospheric constraints can be achieved through advanced
exoplanet transit spectroscopic observations. These consist of
time-series observations of planetary systems as a planet passes in
front of (transit) or behind (secondary eclipse) its host star, as
viewed from Earth (Figure \ref{fig:fig2}). The difference between the
in-transit and out-of-transit flux reveals the amount of starlight
absorbed by the planet's atmosphere during a transit, or the thermal
light emitted by the planet during an eclipse. Each atmospheric
species leaves a unique spectroscopic absorption fingerprint, enabling
the identification of specific molecular and atomic constituents
(Figure \ref{fig:fig2}, right panel).  Furthermore, different
absorption features probe distinct atmospheric layers—infrared
molecular bands typically trace the lower atmosphere (pressure
$p>10^{-6}$~bar), while ultraviolet atomic lines are more sensitive to
the upper atmosphere ($p < 10^{-6}$~bar). Therefore, a
comprehensive understanding of an exoplanet's atmosphere requires
broad, simultaneous coverage across the electromagnetic spectrum, from
the ultraviolet to the infrared ($\sim$120--1600~nm, Figure \ref{fig:fig3}). Lastly, if
these measurements are also sensitive to polarized light, we would be
able to disentangle the light reflected off a planet from that
directly received from its host star, as well as probe aerosols and
other scattering phenomena in the planetary atmosphere.

All in all, these observations have the potential to yield a direct
view into the formation and evolution processes of exoplanets, and
further characterize their composition and thermal structure, which
ultimately will let us probe for the signatures of a habitable
environment.

\section{Physical Parameters}


\subsection{Constraining the Rate and Composition of the Escaping Atmospheric Species}

Observations in the ultraviolet (UV) offer a unique opportunity to
characterize exoplanet atmospheres: they probe the uppermost layers of
an atmosphere (pressures below $\sim$1 $\upmu$bar), where atmospheric
escape is expected to take place
\citep[e.g.,][]{FossatiEtal2015asslExoAtmObservations,
  DosSantos2023iauPlanetaryWindsOutflows}. By detecting and
characterizing escaping species, we can thus constrain fundamental
properties such as thermal structure, mass loss, compositional
evolution, and interactions with the host star.
This also highlights the need for having high-accuracy
laboratory data in the ultraviolet \citep[see, e.g.,][]{FortneyEtal2019astro2020LabDataNeeds}.  These observations are
particularly valuable for understanding how different classes of
exoplanets retain or lose their atmospheres, thereby informing our
broader understanding of planetary formation and evolution
\citep[e.g.,][]{MoranEtal2023apjGJ486bJWSTwaterOrStar,
  OwenWu20217apjEvaporationValley}. Furthermore, UV observations are
only possible from space-based observatories, as Earth's atmosphere is
opaque to most ultraviolet radiation.

\subsubsection*{Far-ultraviolet observations}

\begin{figure}[t!]
\centering
\includegraphics[width=0.95\linewidth, clip]{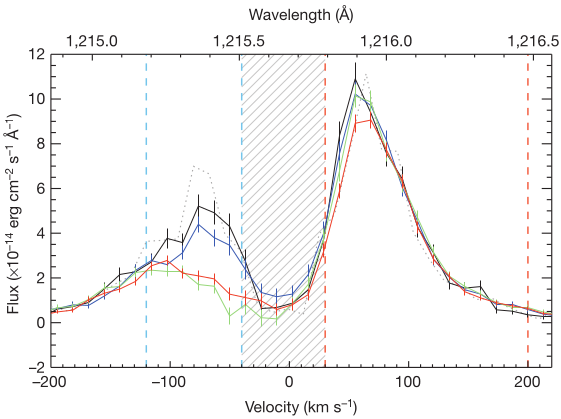}
\includegraphics[width=0.92\linewidth, clip, trim=0 0 0 700]{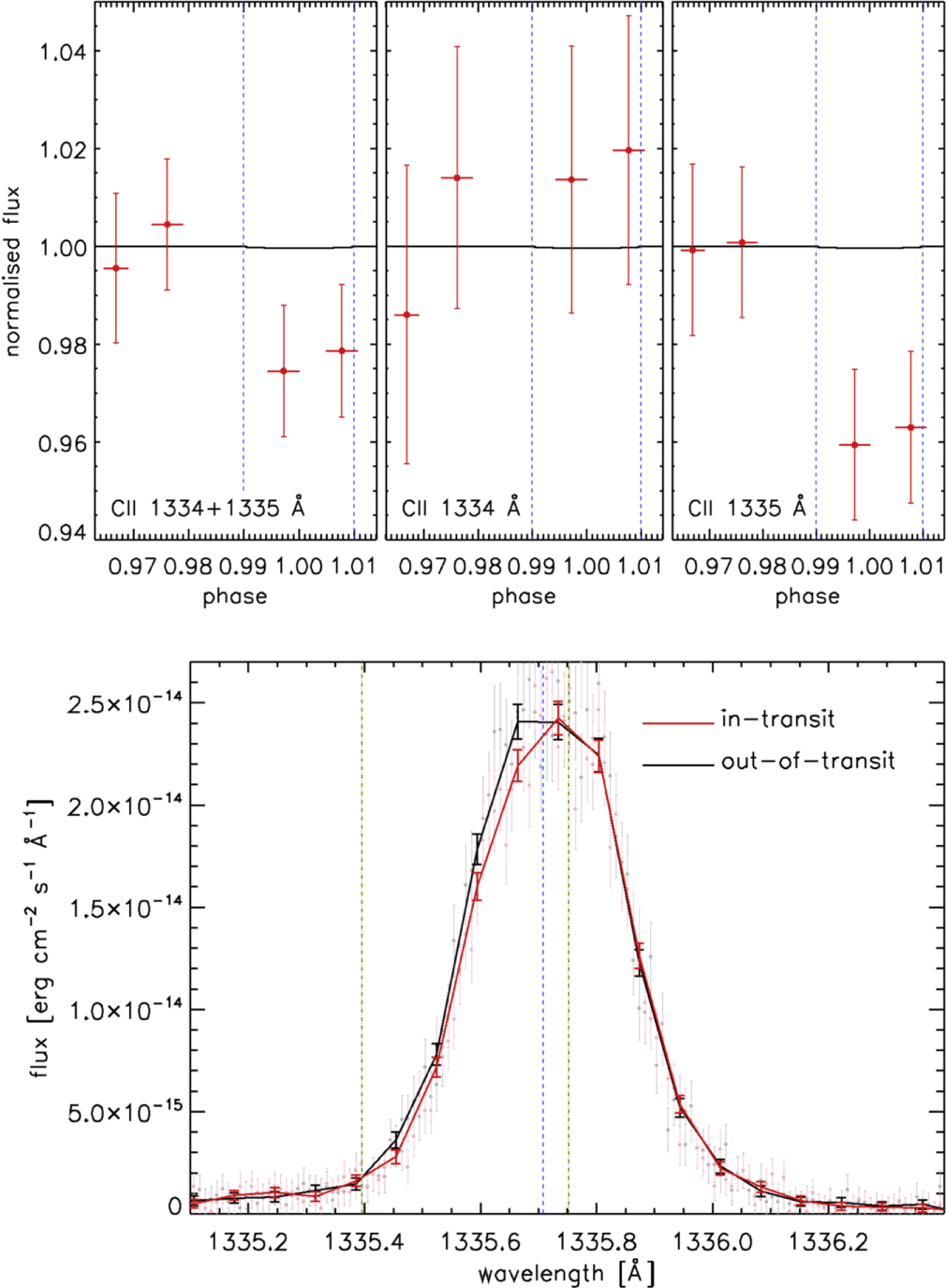}
\caption{Top panel: {\Lyalpha}
  spectra of the M-dwarf GJ436 obtained during the transit of the
  close-in Neptune-like planet GJ 436\,b (red and green lines) compared
  to spectra obtained out-of-transit (black and blue lines). The $\sim$50\%
  absorption in the blue wing of the {\Lyalpha} line indicates the
  presence of a significant amount of neutral hydrogen lost by the
  planet and being blown away (i.e. moving towards us) by the stellar
  wind
  \citep[from][]{EhrenreichEtal2015natGJ436bHydrogenEscape}. Bottom
  panel: in-transit (red) and out-of-transit (black) spectra around
  the position of the \ion{C}{II} 1335 Å doublet for the super-Earth planet $\pi$~Men\,c. The $\sim$7\% absorption in the blue wing of the doublet indicates
  the presence of ionized carbon escaping from the planetary
  atmosphere and moving towards the observer under the action of the
  stellar wind
  \citep[from][]{GarciaMunozEtal2021apjPiMencCarbonEnvelope}.  }
\label{fig:fig4}
\end{figure}

One of the primary applications of FUV observations is the detection
of Lyman-alpha (\Lyalpha) absorption, which traces neutral hydrogen
escaping from an exoplanet's atmosphere. This feature is particularly
useful for detecting and quantifying atmospheric escape, which occurs
when high-energy stellar irradiation heats the upper atmosphere,
leading to the expansion and eventual loss of the lighter elements
\citep{Yelle2004icarGiantPlanetsAeronomy,
  GarciaMunoz2007pssHD209458bAeronomy,
  KoskinenEtal2013icarHD209458bMetalsEscapeI}. The presence of an
extended hydrogen atmosphere, as observed in several hot gas giants
and warm Neptunes, serves as a direct probe of ongoing mass loss
\citep[Figure \ref{fig:fig4}, see
  e.g.,][]{VidalMadjarEtal2003natHD209458bUpperAtmosphere,
  EhrenreichEtal2015natGJ436bHydrogenEscape}. Furthermore, stellar
winds directly interact with the extended upper atmospheres of
exoplanets, modulating their extent, and shaping the geometry of
atmospheric escape
\citep[e.g.,][]{VidottoCleary2020mnrasStellarWindsEscape,
  CarolanEtal2021mnrasStellarWindLyalpha}. Thus, combining
spectroscopic {\Lyalpha} absorption measurements with theoretical
models help constrain the stellar wind properties of the host star,
which significantly influence atmospheric evolution
\citep[e.g.,][]{Wood2004lrspStellarWinds,
  KislyakovaEtal2014sciHD209458bLyAlpha,
  KislyakovaEtal2024natasXrayDetection}.

Also in the FUV band, transit observations at the position of singly
ionized carbon (\ion{C}{II}) and neutral oxygen (\ion{O}{I}) lines enable one to
detect heavier species escaping from exoplanetary atmospheres. The
presence of \ion{C}{II} and/or \ion{O}{I} absorption coming from the upper
atmosphere suggests that planets may be experiencing significant
atmospheric loss. Furthermore, by measuring the relative absorption
strengths at the position of carbon and/or oxygen and hydrogen UV
lines, we can infer the elemental composition of the lower atmosphere
and constrain compositional changes over time, shedding light on the
processes that govern atmospheric retention and loss \citep[Figure \ref{fig:fig4};
  see e.g.,][]{GarciaMunozEtal2020apjPiMencLymanAlpha,
  GarciaMunozEtal2021apjPiMencCarbonEnvelope}. The investigation of
elements heavier than hydrogen and helium in the FUV is particularly
valuable in the characterization of small planets, for which IR
observations may prove too challenging.


\subsubsection*{Near-ultraviolet observations}

\begin{figure*}[t!]
\centering
\includegraphics[width=0.42\linewidth, clip]{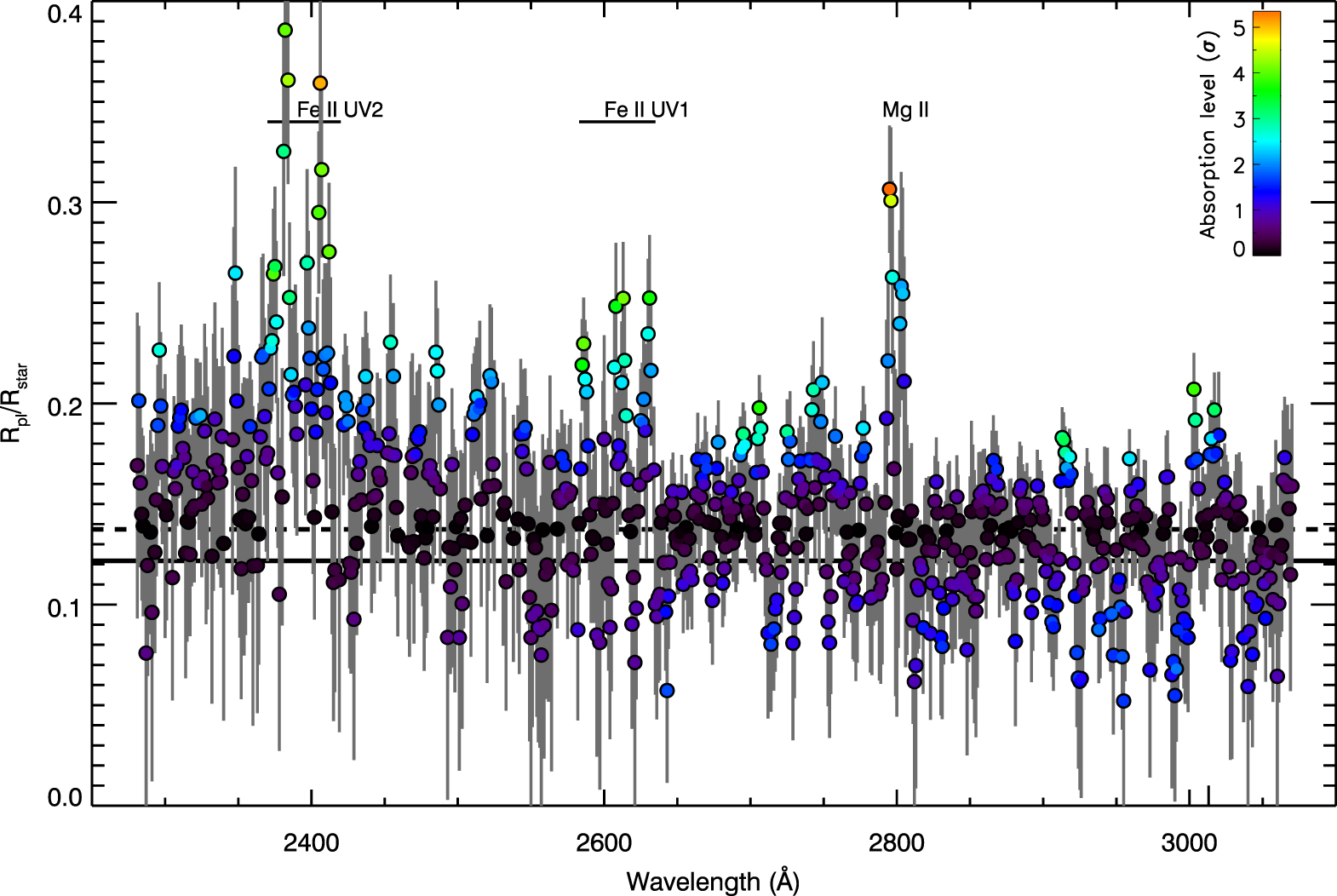}\hfill
\includegraphics[width=0.55\linewidth, clip, trim=0 130 0 0]{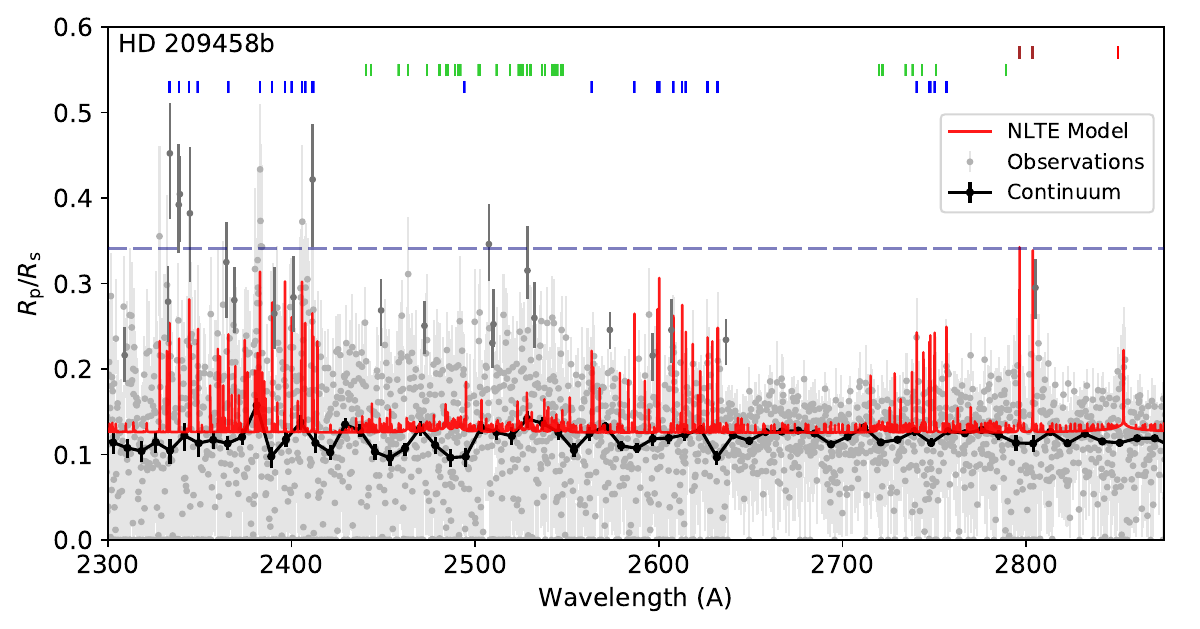}
\caption{NUV transmission spectra obtained with HST/STIS at a
  resolution of $\sim$30,000 for the hot Jupiters WASP-121\,b \citep[left panel;
  from][]{SingEtal2019ajWASP121bTransmissionNUV} and HD~209458\,b \citep[right
    panel; adapted from][]{CubillosEtal2020ajHD209458bNUV}. Both plots
  highlight the position of Fe and Mg lines indicating their detection
  or non-detection, as well as their location with respect to the
  planetary Roche lobe (dashed line). Absorption above the Roche lobe
  at the position of features of a given element indicates that the
  element is entailed in the atmospheric mass loss. If just one of the
  two atoms (Fe or Mg) is escaping from the planet (e.g., HD~209458\,b
  in the right panel for which just Fe is detected) suggests that the
  other one is likely to be trapped in aerosols in the lower
  atmosphere, thus constraining properties (e.g. temperature, mixing
  strength) of atmospheric layers not probed by the NUV observations.}
\label{fig:fig5}
\end{figure*}

NUV observations complement FUV studies by allowing the detection of
heavy metal species in the upper atmospheres of exoplanets. Elements
such as magnesium ({\MgII}), iron (\FeII), and silicon (\SiII) exhibit
strong absorption in this wavelength range, providing constraints on
the chemical diversity of the escaping material and the thermal
structure of the atmosphere \citep[Figure \ref{fig:fig5}, see also e.g.,
][]{FossatiEtal2010apjWASP12bMetals,
  SingEtal2019ajWASP121bTransmissionNUV}. The study of metal ion
absorption features enables us to probe the interaction between
planetary outflows and the stellar wind, helping to refine models of
atmospheric escape and exoplanet-star interactions
\citep[e.g.,][]{HaswellEtal2012apjWASP12bNUV,
  DwivediEtal2019mnrasWASP12bMagnesiumModeling}, and give information
on the efficiency of heating and cooling processes ongoing on all
classes of planets, including Earth and the other solar system planets
\citep[e.g.,][]{SreejithEtal2023apjWASP189bCUTE}.

By analyzing mass-loss rates and atmospheric composition, we can infer
whether a planet has a secondary atmosphere, potentially replenished
through volcanic outgassing or other processes
\citep[e.g.,][]{SeligmanEtal2024apjTidalDissipationMelting,
  BelloArufeEtal2025apjVolcanismL98-59b}. Furthermore, the presence
and composition of aerosols—key indicators of photochemical
processes—can be inferred from UV spectra, providing additional
context for understanding planetary habitability
\citep[e.g.,][]{CubillosEtal2020ajHD209458bNUV,
  CubillosEtal2023aaHD189733bSTISnuv,
  LothringerEtal2020apjUVtransmissionSpectra,
  FromontEtal2024apjAtmosphericEscapeL98-59}.

\subsubsection*{Infrared and Optical Observations}

\begin{figure}[t!]
\centering
\includegraphics[width=\linewidth, clip, trim=0 880 0 900]{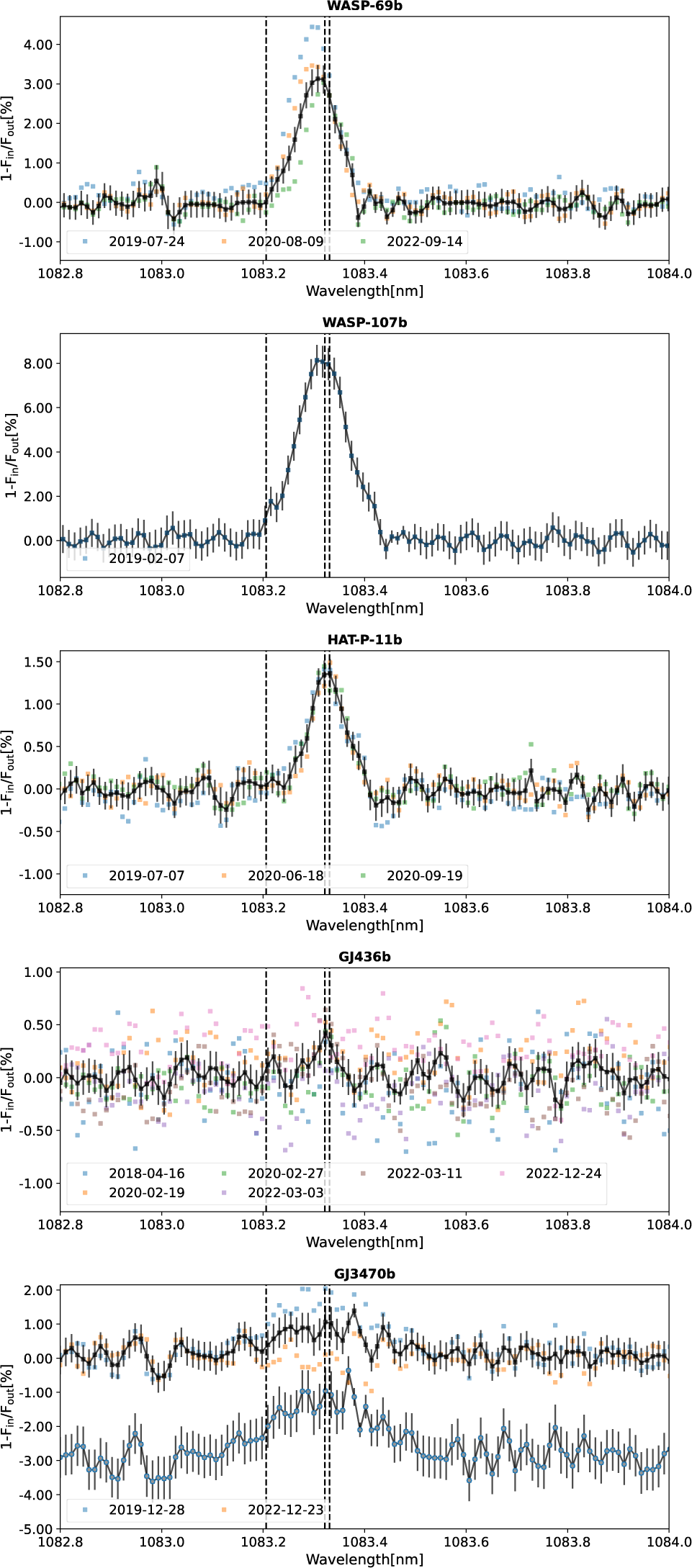}
\caption{High spectral resolution ground-based transmission spectra of
  the
  close-in Neptune-mass planets HAT-P-11b
  \citep[from][]{GuilluyEtal2024aaMKheliumsurvey} at the position of
  the {\helium} metastable triplet, marked by the vertical dashed
  lines. The depth of the absorption is proportional to the strength
  of the atmospheric mass-loss, while the wavelength displacement of
  the absorption relative to the laboratory wavelength of the lines
  probes mainly the interaction with the stellar wind.}
\label{fig:fig6}
\end{figure}

Another way for studying upper atmospheres, and therefore atmospheric
escape is looking for exoplanet atmospheric absorption at the position
of the metastable Helium I triplet, which lies in the near-infrared
\citep[e.g.,][]{SeagerSasselov2000apjExoplanetTransmissionSpectra,
  OklopicHirata2018apjHeliumEscape}. These observations, which are
complementary to UV observations
\citep[e.g.,][]{BerezutskyEtal2024mnrasNeptunesLyalphaHelium}, can
constrain a broad range of stellar (e.g., wind density and velocity)
and planetary (e.g., upper atmospheric temperature and {\helium} abundance)
properties \citep[Figure \ref{fig:fig6}; e.g.,
][]{BiassoniEtal2024aaHeliumModeling,
  SanzForcadaEtal2025aaHeliumExtremeUV,
  BallabioOwen2025mnrasHeliumEscape, McCreeryEtal2025apjHeliumEscape}.

So far, most {\helium} observations have been carried out from
ground-based observatories, and thus they are affected by telluric
contamination \citep[e.g.,][]{AllartEtal2018sciHATP11bHelium,
  SalzEtal2018aaHeHD189733b,
  GuilluyEtal2024aaMKheliumsurvey}. Correcting for telluric
contamination is the single-most challenging aspect of exoplanet
characterization from ground-based telescopes. Unless the observed
system has a significant doppler shift, the Earth atmospheric lines,
whose strength varies across an observing night, will overlap with the
exoplanet lines. In spatially unresolved observations, telluric lines
modulate the spectrum of the host star, which are at least 1000 times
brighter than the atmospheric signature of an exoplanet. Therefore,
even a small inaccuracy in correcting telluric lines leaves an
imprinted residual, which is orders of magnitude stronger than the
exoplanet signature.

Another problem is that the {\helium} lines form in the host star as
well, and they are highly reactive to stellar activity variations
\citep[e.g.,][]{GuilluyEtal2020aaGianoHD189733bHelium}. If the host is
an active star, one has to disentangle stellar and planetary
signals. This requires monitoring of the stellar activity
simultaneously with the planetary transit observations.  This can be
accomplished by observing stellar lines in the optical band, such as H
alpha, \ion{Na}{ID}, and \ion{Ca}{II} H\&K, but they are not optimal and significantly
more and secure information would come from the simultaneous
monitoring of UV lines. Thus, simultaneous {\helium} and UV monitoring
would allow us to properly account for stellar variability in the
interpretation of the observations. Finally, the addition of
spectropolarimetry would enable one to detect and measure exoplanetary
magnetic fields down to strengths of about 1--10 G for polarimetric
sensitivities of $10^{-5}$--$10^{-4}$
\citep{OklopicEtal2020apjHeliumMagneticFields}. The detection and
measurement of exoplanetary magnetic fields would significantly
improve our understanding of their impact on atmospheric mass-loss
rates, and thus on atmospheric evolution as a whole
\citep[e.g.,][]{CarolanEtal2021mnrasMagneticFieldsEscape,
  HazraEtal2025mnrasCMEexoplanetInteraction}.

\subsection{Constraining the Composition and Thermal Structure of the Lower
Atmospheres (High-resolution Cross-correlation)}

\begin{figure}[t!]
\centering
\includegraphics[width=\linewidth, clip]{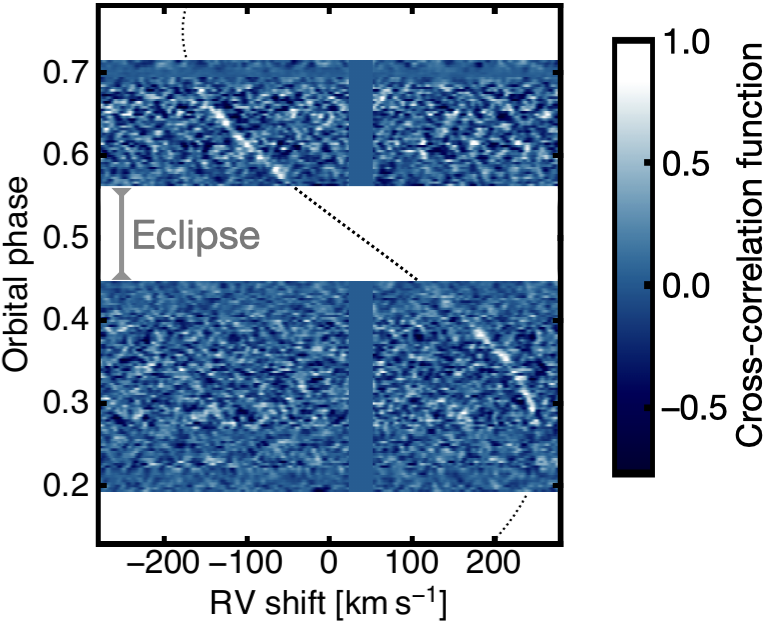}
\caption{Cross-correlation signature of the ultra hot-Jupiter
  WASP-121b as a function of orbital phase and radial-velocity (RV)
  shift, collected from 8 ESPRESSO observing nights \citep[0.38--0.78
    µm,][]{PelletierEtal2025ajWASP121bCRIRESplus}. The peak of the
  cross correlation function (white trace) follows the expected
  orbital RV trace of the planet (black dotted line), indicating that
  the Fe and Ni-refractories included in the template model are
  present in the data.}
\label{fig:fig7}
\end{figure}

In the near-infrared, molecular species exhibit millions to billions
of ro-vibrational transitions, which at the highest resolving power
produce thousands of spectral lines that provide critical insights
into the properties of planetary primary and secondary
atmospheres. However, individual spectral lines are typically too weak
to be detected, even when using the largest ground-based
telescopes. High-resolution cross-correlation spectroscopy (HRCCS) has
emerged as a powerful technique to overcome this limitation by
leveraging the collective signal of all known spectral lines of a
given molecule. This is achieved by cross-correlating the observed
planetary spectrum with a theoretical model template, effectively
boosting the planetary signal and enabling molecular detections
\citep[Figure \ref{fig:fig7}; see e.g.,][]{SnellenEtal2010natHD209458b,
  Birkby2018haexHighResolutionReview}.

One of the key advantages of HRCCS is its ability to distinguish
planetary signals from stellar contamination. At high spectral
resolution ($R \gtrsim 60,000$), stellar and planetary lines exhibit
distinct Doppler shifts due to their differing radial velocities over
the course of the observation. This allows stellar features to be
effectively disentangled from the planetary lines
\citep{BrogiEtal2012natTauboo}. The capacity to separate stellar and
planetary lines further allow us to apply this technique to
non-transiting systems, this time by looking at the thermal emission
from a planet as its emission lines shift over time along the planet's
orbital motion around its host star
\citep[e.g.,][]{BrogiEtal2012natTauboo, RodlerEtal2012apjTauBoo, 
  deKokEtal2013aaCarbonMonoxideHD189733b,
  BirkbyEtal2013mnrasHD189733bHighRes}. Therefore, a stable
high-resolution UVOIR spectrograph on-board HWO would have the ability
to detect and measure atmospheric features and dynamics of long-period
habitable-zone planets.  This would open the possibility of detecting
and eventually measuring the abundance of potential biosignature
molecules, such as oxygen and ozone.

HRCCS in the infrared predominantly probes the deeper layers of the
atmosphere where molecular absorption features form (a consequence of
targeting molecular ro-vibrational lines with relatively low opacity),
allowing constraints on both the chemical composition and thermal
structure of the lower atmosphere
\citep[$p<10^{-2}$~bar,][]{GandhiEtal2020mnrasHiresMolecularCrossSections}. However,
ground-based observations face two primary challenges. First, telluric
contamination from Earth's atmosphere introduces absorption features
that can interfere with planetary signals, making it essential to
robustly differentiate planetary spectral features from terrestrial
ones \citep[e.g.,][]{AllartEtal2017aaHD189733bWaterVapor}. Second, in
contrast to low-resolution observations, HRCCS data are
self-calibrated such that the broadband and time variations of the ﬂux
at each wavelength are divided out of the data
\citep{BrogiLine2019ajHiResRetrievals}. This results in the loss of a
reliable continuum information, limiting constraints on aerosols and
clouds, which are typically inferred from spectral slopes and absolute
flux measurements
\citep[e.g.,][]{kitzmannHeng2018mnrasCondensateOpticalProperties,
  LineParmentier2016apjPartialClouds}.

Despite these challenges, HRCCS has revolutionized the detection of
molecular species such as water, carbon monoxide, and methane in
exoplanet atmospheres, particularly in hot Jupiters. Space
observations at high resolution, not being affected by tellurics, will
enable more straightforward analysis techniques where the stellar
spectrum can be removed by creating a template from the out-of-transit
spectrum. This will further enable HRS observations of cooler
temperate exoplanets (larger orbits) who will exhibit slower RV
Doppler shifts with time than hot Jupiters. Furthermore, the large HWO
collecting area would enable us to probe the atmospheres of smaller
planets, getting us closer to probe the environment of rocky habitable
zone worlds.


{
\begin{table*}[t!]
\centering
\small
\caption[Performance Goals]{Exoplanet constraints from upper, lower, and combined atmospheric regions.  Benchmarks for HWO configurations (GB: ground-based high-resolution cross correlation)}
\label{tab:performance}
\begin{tabular}{|p{2.625cm}|p{3.2cm}|p{3.0cm}|p{3.0cm}|p{3.0cm}|}
\noalign{\smallskip}
\hline
\noalign{\smallskip}
{{\bf Physical Parameter}} & {{\bf State of the Art}} & {{\bf Incremental Progress}} & {{\bf Substantial Progress}} & {{\bf Major Progress}} \\
\noalign{\smallskip}
\hline
\noalign{\smallskip}
Upper-atmosphere escape (FUV) &
{\Lyalpha} in $\sim$5 Jupiter planets (HST); \ion{C}{I} or \ion{O}{II} in $\sim$5 Neptune to Jupiter-like planets (HST) &
{\Lyalpha} and metal escape-rates in 10 sub-Neptune to Jupiter planets &
{\Lyalpha} and metal escape-rates in 50 super-Earth to Jupiter planets &
{\Lyalpha} and metal escape-rates in 100 Earth to Jupiter planets \\
\hline
Upper-atmosphere escape (NUV) &
Neutral and ionized Fe and Mg detection/non-detection in $\sim$5 Jupiter-like planets (HST, CUTE) &
Fe and Mg escape-rates in 10 Neptune to Jupiter-like planets &
Fe and Mg escape-rates in 50 Super-Earth to Jupiter-like planets &
Fe and Mg escape-rates in 100 Earth to Jupiter-like planets \\
\hline
Upper-atmosphere escape (NIR) &
{\helium} in $\sim$40 Neptune to Jupiter-like planets (GB, HST, JWST) &
{\helium} constraint with concurrent optical monitoring of stellar activity (before and after transit) in 50 Neptune to Jupiter-like planets &
{\helium} constraint with concurrent UV monitoring of stellar activity (before, after, and/or during transits) in 100 super-Earth to Jupiter-like planets &
{\helium} constraint with simultaneous UV monitoring of stellar activity (before, after, and during transits) in 200 Earth to Jupiter-like planets \\
\hline
Lower-atmosphere composition &
Detection of \ch{H2O}, CO, \ch{CO2}, \ch{CH4}, \ch{NH3}, \ch{SO2},
\ch{H2S}, \ch{SiO}, Na, K (at $\sim$1 dex precision or worse) in $\sim$60--80 Neptune-
to Jupiter-like planets (HST, JWST, and GB). Typically no more than 1--3 species per target &
Abundance of volatiles (\ch{H2O}, \ch{CO2}, \ch{CH4}) plus refractories (Na, K, \ch{SO2}) to 0.3 dex in sub-Neptune planets &
Inventory of multiple volatile plus refractory abundances (or non-detection) Earth to Jupiter-like planets &
Full inventory of volatile plus refractory abundances (or non-detection) in Earth to Jupiter-like planets \\
\hline
Lower-atmosphere continuum &
Rayleigh-like hazes and cloud condensates in $\sim$50 planets (HST, JWST). Often degenerate with high mean molecular weight atmospheres &
Condensate detection disentangled from high mean molecular weight atmospheres &
Unambiguous detection of condensate precursors (e.g., soots, MnS, \ch{MgSiO3}) &
Unambiguous detection of condensate precursors plus spatial constraints (e.g., east/west limbs, patchyness fraction) \\
\hline
Spectropolarimetry & Stokes I & Stokes I and Q 0.01\% &
Stokes I, Q, U, and V 0.01\% &
Stokes I, Q, U, and V 0.001\% \\
\hline
Combined upper and lower atmosphere &
2 hot-Jupiter planets, from non simultaneous observations &
5 Neptune to Jupiter-like planets, non simultaneous &
10 Earth to Jupiter-like planets, near simultaneous &
50 Earth to Jupiter-like planets, simultaneous \\
\noalign{\smallskip}
\hline
\end{tabular}
\end{table*}
}

\subsection{Spectropolarimetry}

Reflected starlight provides insight into the envelope of a planet
that cannot be obtained with any other technique. A planet's
brightness in reflected starlight, and its evolution with the
star-planet-observer phase angle, depend on, for example, the single
scattering albedo of the dominant scattering particles, which is
linked to the particles' composition and size. This fundamental idea
has been used to characterize the clouds of a few hot Jupiters for
which broadband phase curves were measured with Kepler or TESS
\citep{DemoryEtal2013apjKepler7bReflectiveClouds,
  GarciaMunozIsaak2015pnasPhaseCurveClouds}.  HRCCS offers an
alternative to how reflected-starlight phase curves can be measured.
The light reflected off the planet is a Doppler-shifted copy of the
stellar spectrum, with additional distortions introduced by the gases
and aerosols in the planet's atmosphere.  The application of HRCCS to
reflected-starlight measurements is a theme where significant
development is expected in the immediate future, possibly in
combination with high-contrast imaging systems
\citep{SnellenEtal2015aaHighDispersionHighContrast}. The technique
will be crucial in the characterization of non-transiting, temperate
exoplanets for which thermal emission measurements are
impractical. The technique becomes especially powerful if the spectra
can be combined to determine the Q and U elements of the polarization
Stokes vector \citep{GarciaMunoz2018apjPolarimetryMapping}. Indeed,
polarization is more sensitive than brightness-only measurements to
scattering phenomena such as the rainbow or the glory
\citep{Bailey2007asbioExoplanetPolatization}. These phenomena can be
used to assess whether the clouds are in the liquid phase, as
demonstrated in the study of Venus
\citep{HansenHovenier1974jatisVenusPolarization}. The available
simulations of HRCCS including polarimetry
\citep{GarciaMunoz2018apjPolarimetryMapping} for the characterization
of close-in exoplanets around bright stars show that the technique
presents unique advantages. Indeed, and unlike broadband polarimetry,
the application of HRCCS polarimetry facilitates the removal of the
polarization introduced by the interstellar medium or the host star,
signals that can be comparable in magnitude to the polarization of the
planet.

\section{Description of Observations}


The full and simultaneous characterization of an exoplanetary (upper
and lower) atmosphere has never been done before. Therefore, we
propose to carry out a survey of order 50 transiting planets covering
a wide range of system parameters, from gas giants to Earths, over a
wide range of orbits and stellar spectral types. Thanks to TESS,
CHEOPS, and soon PLATO the list of long-period (from $\sim$40 to 100+
days) transiting planets orbiting bright stars is rapidly
growing. Thanks to HWO's coronagraphic capabilities, the list of
long-period planets orbiting nearby stars will further grow. These
long-period planets, particularly those in the Earth to sub-Neptune
regime \citep[e.g.,][]{DelrezEtal2021natasCHEOPSnuLupd}, will have
priority, because their irradiation level is similar to that of the
solar system planets, boosting comparative planetology. The primary
goal of PLATO, to be launched at the end of 2026, is to look for
transiting low-mass planets orbiting in the habitable zone of Sun-like
stars. As such, these planets will become golden targets for HWO
atmospheric characterization observations through both the coronagraph
and transmission/emission spectroscopy.

The first years of JWST exoplanet science have brought exciting and
unprecedented advancements for atmospheric characterization. These
include the first unambiguous spectroscopic detection of species such
as \ch{CO2} or \ch{SO2}
\citep[e.g.,][]{ERSteam2023natCarbonDioxideWASP39b, TsaiEtal2023natWASP39bPhotochemistry,
  WelbanksEtal2024natWASP107bJWSTtransits}, the first tentative
detection of molecules like \ch{H2S} and \ch{CS2}
\citep[e.g.,][]{FuEtal2024natasHD189733bH2S,
  BennekeEtal2024arxivTOI270dMiscible}, and evidence for non-gray
condensates \citep[e.g.,][]{GrantEtal2023apjWASP17bQuartzClouds,
  DyrekEtal2024natWASP107bTransitMIRI}. Furthermore, JWST has
demonstrated the capability to spatially map some of these atmospheric
properties
\citep[e.g.,][]{EspinozaEtal2024natWASP39bInhomogeneousTerminators,
  ChallenerEtal2024apjWASP43bMultiDimEmission}.  However, JWST has
also highlighted the challenges involved in characterizing exoplanet
atmospheres. Most spectroscopic and spatially resolved detections to
date cover Jupiter- to Neptune-sized planets, with only a few
exceptions \citep[e.g., TOI-270\,d or
  K2-18\,b;][]{HolmbergMadhusudhanEtal2024aaTOI270dJWSThycean,
  MadhusudhanEtal2023apjK2-18bJWSTcarbonMolecules,
  SchmidtEtal2025arxivK2-18bReanalysis}. In contrast, observations of
smaller planets have often yielded featureless spectra, limiting their
physical interpretation to bulk properties. Instrumental systematic
depth offsets \citep{CarterEtal2024natasDataSynthesisWASP39b} and
stellar activity have emerged as a major obstacles to unambiguously
identify planetary features
\citep[e.g.,][]{MoranEtal2023apjGJ486bJWSTwaterOrStar,
  RathckeEtal2025apjTRAPPIST1stellarCorrection}, complicating efforts
aiming to build up signal-to-noise through multiple observations. This
only highlights the potential of a future observatory like HWO to
characterize exoplanet atmospheres, while reinforcing the need for
observations with broad, ideally simultaneous, UV-to-NIR coverage to
constrain both stellar and planetary properties at the same time.

The almost totality of the atmospheric characterization observations
conducted to date focuses on one specific wavelength band (e.g., NUV,
IR), or even on specific features (e.g., {\Lyalpha},
\ion{C}{II}). Furthermore, high-resolution observations have been
limited to ground-based facilities.  This carries severe limitations
that would be overcome by having on-board HWO a stable,
high-resolution ($R>60,000$) spectropolarimeter (e.g., Pollux) capable
of covering simultaneously from the {\Lyalpha} line (120 nm) to
1.6--1.7 µm (i.e. what currently covered by WFC3 on-board HST).

\begin{itemize}
\item An instrument with these capabilities on-board HWO would
  uniquely enable us to observe, and thus simultaneously constrain,
  the properties of an entire atmosphere, from the region close to the
  continuum (bar to mbar level) to the uppermost layers possibly
  escaping to space (above the nbar level). This kind of observation
  is currently impossible, and will remain so until the advent of HWO,
  because of programmatic (difficulty in coordinating ground- and
  space-based time constrained observations) and instrumental
  (unavailability of a stable instrument covering simultaneously the
  FUV and NUV bands) reasons. Any dataset obtained with such an
  instrument would have a far reaching legacy.

\item The simultaneous coverage of such a broad wavelength range would
  remove uncertainties in the interpretation of the observations
  related to possible time variability connected to changes in stellar
  activity and/or possible planetary atmospheric temporal
  variability. Therefore, such an instrument would effectively pave
  the way to directly studying, for example, the impact of stellar
  high-energy emission, probed by UV lines, on photoionization and
  photochemistry. Furthermore, these observations would help
  understanding the origin of the discrepancies often obtained between
  ground- and space-based observations
  \citep[e.g.,][]{GiacobbeEtal2021natSixMoleculesCNO,
    XueEtal2024apjHD209458bJWSTtransits}.

\item The stability in flux and wavelength of the instrument, together
  with the high spectral resolution, would enable one to employ
  cross-correlation analysis techniques on space data. This would have
  the great advantage over currently performed ground-based
  observations of not losing the continuum absorption of the planetary
  atmosphere, which carries critical information on atmospheric
  aerosols, removing degeneracies currently present in the
  interpretation of medium- (i.e., space) or high-resolution (i.e.,
  ground) observations.

\item The key further advantage of a space-based stable
  high-resolution spectrograph is the lack of telluric contamination,
  which is still one of the main hurdles in the analysis of
  high-resolution ground-based transmission/emission spectroscopy
  observations.

\item This advantage applies also to observations of reflected
  starlight. In particular, if the observations aim to constrain the
  polarization of the planet, going to space removes at once the
  time-varying, and thus largely unpredictable, polarization
  introduced by the terrestrial atmosphere.

\item Finally, the simultaneous UV and high-resolution coverage would
  open the window to a completely new set of information currently
  completely unexplored. First, it provides access to molecules that
  are either (i) of difficult access as a result of the weakness of
  the absorption bands (e.g., \ch{O2}, \ch{O3}) or (ii) accessible
  only at much longer wavelengths unavailable as part of an UVOIR
  observatory (e.g., \ch{CO2}) or (iii) inaccessible even at longer
  wavelengths (e.g., \ch{NO2}). Second, by combining the information coming
  from bands/lines of the same molecules resulting from
  rotational/vibrations transitions (nIR) and electronic transition
  (UV), and thus rising from different parts of the atmosphere, the
  observations would constrain atmospheric abundance profiles and
  test, for example, how deep photochemistry impacts a planetary
  atmosphere.

\end{itemize}

Location of the mission far from Earth, e.g. at L2, ensures that
critical UV lines, such as {\Lyalpha} and \ion{O}{I} triplet are not affected by
geocoronal emission, enabling to more easily reconstruct the
unabsorbed stellar {\Lyalpha} profile and measure \ion{O}{I} planetary
absorption. Furthermore, the FUV observations will require low
detector noise, possibly better than what is currently present for
HST/COS.

The polarimetric capability requires high wavelength stability, to at
least a tenth of a pixel. To enable transit/eclipse observations of
long-period planets (e.g., 1 year) and phase curve observations of
close-in planets, such stability has to be ensured over at least 30
hours. The L2 location of HWO helps reach these requirements.

\begin{table*}[ht!]
\centering
\small
\caption[Observation Requirements]{Requirements for \textbf{combined UV, optical, and NIR} observations of exoplanet atmospheres and escape (GB: ground-based high-resolution telescopes; OIR: optical plus NIR wavelengths).}
\label{tab:obsreq_oir}
\begin{tabular}{|l|c|c|c|c|}
\noalign{\smallskip}
\hline
\noalign{\smallskip}
{\bf Observation} & {\bf State of the Art} & {\bf Incremental Progress} & {\bf Substantial Progress} & {\bf Major Progress} \\
\noalign{\smallskip}
\hline
\noalign{\smallskip}
Spectral range (nm) & \makecell{81--360 (UV, HST) \\ 350--2500 (OIR, GB)} &  150--1100 &  120--1700 &  100--2500 \\
\hline
Mirror size (m) & \makecell{2.5 (UV, HST) \\ 8.0 (OIR, ESO/VLT)} &  4.0  &  7.0  &  15.0 \\
\hline
Spectral resolving power & \makecell{$\gtrsim$115,000 \\ (echelle spectrographs)} &  60,000 &  100,000 &  $>$100,000 \\
\hline
Wavelength stability & \makecell{5 km\,s$^{-1}$ (UV)\\ 10 cm\,s$^{-1}$ (optical) \\ 10 m\,s$^{-1}$ (NIR)} &  10 m\,s$^{-1}$ &  1 m\,s$^{-1}$ &  $< 1$ m\,s$^{-1}$ \\
\hline
Stokes parameters & \makecell{I (UV) \\ I, Q, U, V (OIR)}  &  I  &  I and Q  &  I, Q, U, and V \\
\hline
Telluric contamination & \makecell{No (UV)\\ Yes (OIR)} & No & No & No \\
\hline
UV to NIR simultaneity & No &  No &  Yes &  Yes \\
\noalign{\smallskip}
\hline
\end{tabular}
\end{table*}




\end{document}